\documentclass[12pt]{article}

\usepackage{graphics,cite,amssymb,epsfig,float,psfrag}
\usepackage[usenames,dvips]{color}
\usepackage{rotating}

\newcommand{\lsim}{\raisebox{-0.13cm}{~\shortstack{$<$ \\[-0.07cm] $\sim$}}~} 
 
\newcommand{\beq}{\begin{eqnarray}} 
\newcommand{\eeq}{\end{eqnarray}} 

\newcommand{\s}{\\ \vspace*{-4mm}}

\begin{document}
%%%%%%%%%%%%%%%%%%%%%%%%%%%%%%%%%%%%%%%%%%%%%%%%%%%%%%%%%%%%%%%%%%%%%%%%%%%%%%%

\vspace{1cm}

\hfill CERN--PH--TH/2010--315

\hfill  LPT--ORSAY--10--107

\vspace*{1.5cm}

\begin{center}

{\large\bf The Tevatron Higgs exclusion limits and theoretical}

\vspace*{1mm} 

{\large\bf  uncertainties: a critical appraisal}

\vspace*{.5cm}

{\large J. Baglio$^1$, A. Djouadi$^{1,2}$, S. Ferrag$^{3}$, 
R.M. Godbole$^{2,4}$} 

\vspace*{.8cm}

$^1$ Laboratoire de Physique Th\'eorique, U. Paris--Sud and  CNRS,  F--91405
Orsay, France.

$^2$ Theory Unit, Department of Physics, CERN, CH-1211 Geneva 23, 
Switzerland.

$^3$ Dpt. of Physics and Astronomy, U. of Glasgow, G12 8QQ Glasgow, United
Kingdom.

$^4$ Center for High Energy Physics, Indian Institute of Science, Bangalore 560 012, 
India.

\end{center}

\vspace*{1cm}

%------------------------------------------------------------------------------
\begin{abstract} 

We examine  the exclusion limits set by the CDF and D0 experiments on the 
Standard Model Higgs boson mass from their searches at the Tevatron in the light
of large theoretical uncertainties on the signal and background cross sections.
We show that when these uncertainties are consistently taken into account, the
sensitivity   of the  experiments becomes significantly lower and the currently
excluded mass range $M_H\!=\!158$--175 GeV could be entirely reopened. The  
necessary luminosity required to recover the current sensitivity is found to be
a factor of two higher than the present one. 

%------------------------------------------------------------------------------
\end{abstract} 

\newpage

With its successful operation in the last years, the Fermilab Tevatron  $p \bar
p$ collider has now collected a substantial amount of integrated luminosity
which allows the CDF and D0 experiments to be sensitive to the Higgs particle,
the remnant of the mechanism that breaks the electroweak gauge symmetry  of the
Standard Model (SM) and is at  the origin of elementary particle masses
\cite{Higgs,Review}. \s

At the Tevatron, the main search channel for the SM Higgs boson is the top and
bottom quark loop mediated  gluon--gluon fusion mechanism $gg\to H$ with the
Higgs boson decaying into $WW$ pairs which lead to the clean $\ell \nu \ell \bar
\nu$  final states with $\ell\!=\!e,\mu$. The subleading Higgs--strahlung
processes $q \bar q\! \to\! WH,ZH$ add a little to the sensitivity, in
particular at low Higgs masses.  Strong constraints  beyond  the well
established LEP bounds \cite{LEP-Higgs}  have been recently set by the CDF and
D0 collaborations  on the Higgs mass and the range  $M_{H}\!=\!158$--175 GeV 
has been excluded at the 95\% confidence level (CL) \cite{Tevatron}. \s

Nevertheless, this exclusion limit relies crucially on the theoretical
predictions for the cross sections of both  the Higgs signal and  the relevant
SM backgrounds which, as  is well known, are affected by significant
uncertainties. In a recent study \cite{Hpaper0}, it has been re-emphasized that
this is indeed the case for the main Higgs search channel at the Tevatron: 
adding all sources of theoretical  uncertainties in a consistent manner, one
obtains an overall uncertainty of  about $\pm 40\%$ on the $gg\! \to \! H \! \to
\! \ell \nu \ell \bar \nu$ signal\footnote{There are also  uncertainties  on the
Higgs decay  branching ratios, but they are very small in the excluded $M_H$
range; see Ref.~\cite{Note-BR}.}. This  is much larger than the uncertainty
assumed in  the CDF/D0 analysis, i.e. 10\% for D0 and  20\% for  CDF, thus 
casting some doubts on the resulting exclusion limit. \s

In this letter, we confront the Tevatron exclusion Higgs limit with the 
theoretical uncertainties that affect the signal and background rates. We show
that when they are included, the sensitivity  of the the CDF/D0 experiments is
significantly lower than the currently quoted one.  We estimate the necessary
luminosity that is required to recover the current sensitivities and find that
it should be higher than the present luminosity by a factor up to  two. \s

We begin our investigation by summarizing the impact of the theoretical
uncertainties on the $gg \to H$ signal cross section which has a threefold
problem. First, the perturbative QCD corrections to the cross section turned out
to be extremely large: the  $K$--factor defined as the ratio of the higher order
to the leading order  (LO) \cite{ggH-LO} cross sections,  is about a factor of
two at next-to-leading order  (NLO) \cite{ggH-NLO} and about a factor of three
at the next-to-next-to-leading  order (NNLO) \cite{ggH-NNLO}.  It is clear that 
it is this exceptionally large $K$--factor which  allows a sensitivity to the
Higgs boson at the Tevatron with the presently collected data. Nevertheless, the
$K$--factor is so large that one may question the reliability of the
perturbative series. As a corollary,   the  possibility of  still large  higher 
order contributions beyond NNLO cannot be excluded. \s

It has become customary to estimate the effects of these yet uncalculated higher
order contributions from the variation of the cross section  with the
(renormalisation $\mu_R$ and  factorisation $\mu_F$) scale at which the process
is evaluated.  Starting from a median scale $\mu_0$ which  is taken to be
$\mu_R\!=\!\mu_F\!=\!\mu_0\!=\!\frac12 M_H$ in the $gg\! \to\! H$ process, the
current convention is to vary these two scales within the range $\mu_0/\kappa
\!\le \!\mu_R, \mu_F \!\le \! \kappa \mu_0$ with the constant factor chosen to
be $\kappa\!=\!2$.  However, as the QCD corrections are so large in the present
case, it is wise to extend the domain of scale variation and adopt instead a
value  $\kappa\!=\!3$. This is the choice made in Ref.~\cite{Hpaper0} which
resulted in an ${\cal O}( 20\%)$ scale uncertainty\footnote{See also 
Ref.~\cite{Scale-H0j} for another reason to increase the scale uncertainty to
20\%.} on   $\sigma^{\rm NNLO}_{gg\! \to\! H}$ . \s

Another problem that is specific to the $gg \to H$ process is that, already at
LO, it occurs at the one--loop level with the additional complication of having
to account for the finite mass of the  loop particle.  This renders the NLO
calculation extremely complicated and the NNLO calculation a formidable  task.
Luckily, one can work in an effective field theory (EFT) approach in which the
heavy loop particles are integrated out, making the calculation of the
contributions beyond NLO possible. While this approach is justified for the
dominant top quark contribution for $M_H\! \lsim\! 2m_t$, it is not valid for
the $b$-quark loop and for those involving the electroweak gauge bosons
\cite{ggH-EW}. The uncertainties induced by the use of the EFT approach at NNLO
are estimated to be  of ${\cal O}(5\%)$ \cite{Hpaper0}. \s

A third problem is due to the presently not satisfactory  determination of the
parton distribution  functions (PDFs). Indeed,  in this process which is
initiated by $gg$ fusion, the gluon  densities are poorly constrained, in
particular in the high Bjorken--$x$  regime which is relevant for Higgs
production at the Tevatron.   Furthermore, since the $gg\!\to\!H$ cross section
is proportional to $\alpha_s^2$ at LO and receives large contributions  at
${\cal O}(\geq \alpha_s^3)$, a small change of $\alpha_s$ leads to a large
variation of $\sigma^{\rm NNLO}_{gg \to H}$. Related  to that  is  the
significant difference between the world average $\alpha_s$ value and the one
from deep-inelastic  scattering (DIS) data used in the PDFs \cite{PDG}. \s

Modern PDF sets provide a method to estimate these uncertainties by allowing  a
1$\sigma$ (or more) excursion of the experimental data that are used to perform
the global fits. In addition, the MSTW collaboration \cite{PDF-MSTW} provides a
scheme that allows for a combined evaluation of the PDF uncertainties and the
(experimental and theoretical) ones on $\alpha_s$. In Ref.~\cite{Hpaper0}, the 
combined 90\% CL PDF+$\Delta^{\rm exp}\alpha_s+\Delta^{\rm th}\alpha_s$
uncertainty on  $\sigma_{gg\to H}^{\rm NNLO}$ at the Tevatron, was found to be
of order 15\%.  However, this (Hessian) method does not account for the
theoretical assumptions that enter into the parametrization of the PDFs.  A way
to access this theoretical uncertainty is to compare the results for the central
values of the cross section with the best--fit PDFs when using  different
parameterizations. \s

\begin{figure}[!h]
\begin{center}
\vspace*{1mm}
\epsfig{file=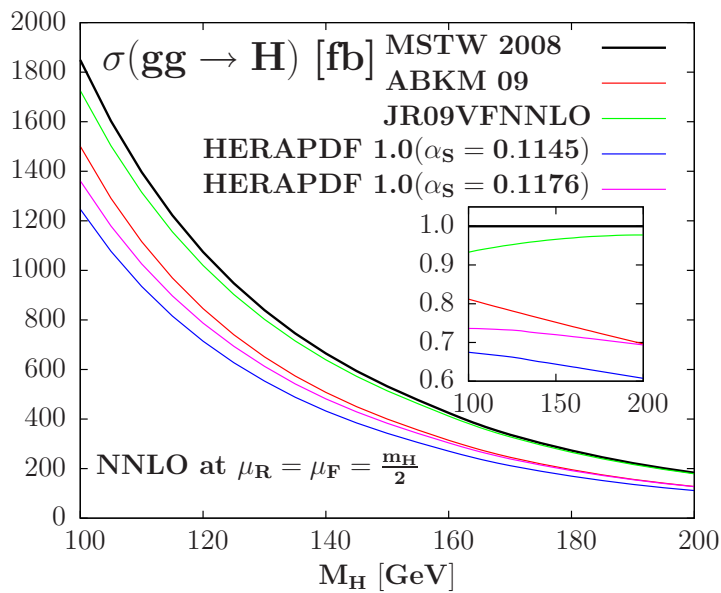,scale=0.89} 
\end{center}
\vspace*{-5mm}
\caption[]{The $gg\to H$ cross section as a function of $M_H$ when the four 
NNLO PDF sets, MSTW, ABKM, JR and HERAPDF, are used. In the inserts, shown
are the deviations with respect to the central MSTW value.}
\vspace*{-2mm}
\label{PDFs}
\end{figure}

In  Fig.~1, displayed are the values of $\sigma^{\rm NNLO}_{gg\to H}$  obtained
when using the gluon  densities that are predicted by the four PDF 
sets\footnote{We consider only NNLO PDFs as we make the choice of  using
partonic cross  sections and PDFs that are consistently taken at the same order
of perturbation theory.}   that have parameterizations at NNLO:  MSTW
\cite{PDF-MSTW},  JR \cite{PDF-JR}, ABKM \cite{PDF-ABKM} and HERAPDF
\cite{PDF-HERA}.  In the later case, two sets are provided: one with an
$\alpha_s$  value that is close to that of MSTW and another one with  the
$\alpha_s$   that is obtained using DIS data alone. As can be seen, there is a
very large spread  in the four predictions,  in particular at large $M_H$ values
where the poorly constrained gluon densities at high--$x$ are involved. The
largest rate is obtained with MSTW, but the cross section using the
HERAPDF set\footnote{It is often argued against the HERAPDF scheme, which uses
consistently only  HERA data to determine the flavour  decomposition,  that it
does not use any jet (Tevatron or DIS) data which is in principle important  in
the determination of the gluon  densities. However, HERAPDF describes well not
only the Tevatron jet data but also   the $W,Z$ data. Since this is a prediction
beyond leading order, it has also  the contributions of the gluon  included.
This gives an indirect test that the  gluon densities are predicted in a
satisfactory way. See also Ref.\cite{PDF-all}.} with the small $\alpha_s$ value
is $\approx\! 40\%$ lower for $M_H\!\approx\! 160$ GeV.\s

A related issue, which is of utmost importance,  is the way these various
uncertainties should be combined. The CDF and D0 experiments  simply add in
quadrature the uncertainties from the scale variation and the PDF uncertainties
obtained through the Hessian method (and ignore the smaller EFT uncertainty) and
they obtain an overall uncertainty of order 20\% on the inclusive cross
section.  We believe (see also Ref.~\cite{ADGSW}) that this procedure has no
justification\footnote{There were some responses to the addendum of
Ref.~\cite{Hpaper0} from CDF and D0 on the {\tt tevnphwg.fnal.gov} web site.
While many comments were made on  secondary and/or agreed points, the main issue
(which explains the difference between our results) is the way to combine the
scale and PDF uncertainties, and it was not really addressed.}. Indeed,  the
uncertainties associated  to the PDFs in a given scheme should be viewed as
purely theoretical uncertainties (due to the theoretical assumptions in the
parameterization) despite of the fact that  they are presented as the $1\sigma$
or more departure from the central values of the data included in the PDF fits.
In some sense, they should be  equivalent to the spread that one observes when
comparing different parameterizations of the PDFs. Thus, the PDF uncertainties
should be considered as having no statistical ground (or a flat prior in
statistical language),  and thus, combined linearly with the uncertainties from
the scale variation and the EFT approach, which are pure theoretical errors.
This is the procedure  recommended, for instance, by the LHC Higgs cross section
working group \cite{LHCXS}. Another, almost equivalent, procedure has been
proposed in Ref.~\cite{Hpaper0}:  one applies the combined PDF--$\alpha_s$
uncertainties directly on the maximal/minimal cross sections with respect to
scale variation\footnote{A similar procedure has also been  advocated  in
Ref.~\cite{Note-proc-sum}  for top quark pair production.}, and then adds
linearly the small uncertainty from the  EFT approach.  This  last procedure,
that we have used here,  provides an overall uncertainty  that is similar (but
slightly smaller) to that obtained with  the linear sum of all uncertainties. \s

\begin{figure}[!h]
\begin{center}
\vspace*{1mm}
\epsfig{file=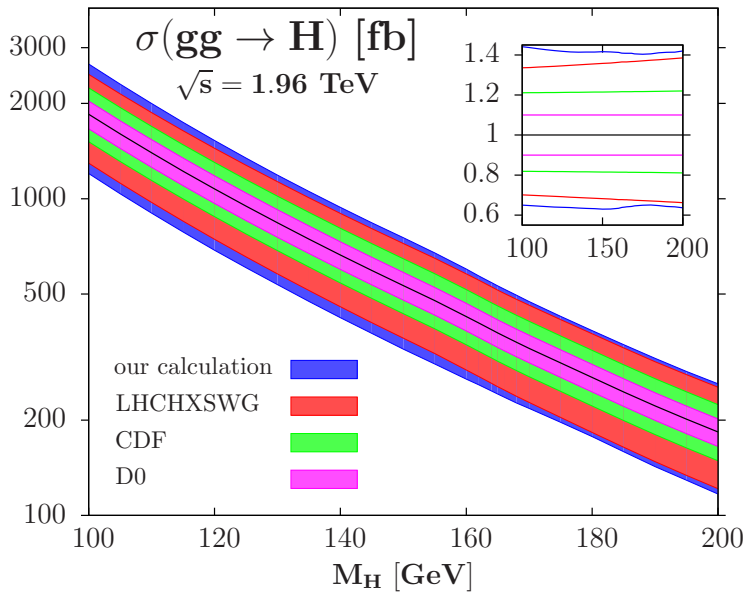,scale=0.92} 
\end{center}
\vspace*{-3mm}
\caption[]{The production cross section $\sigma^{\rm NNLO}_{gg\to H}$  at the 
Tevatron using the MSTW PDFs, with the uncertainty band when all theoretical
uncertainties are added as in Ref.~\cite{Hpaper0} (BD). It is compared 
the uncertainties quoted by the CDF and D0 experiments  \cite{Tevatron} as well
as the uncertainty when the LHC procedure  \cite{LHCXS} is adopted.  In the
insert, the relative size of the   uncertainties compared to the central value
are shown.}
%\vspace*{-1mm}
\label{Total}
\end{figure}

The overall theoretical uncertainty on $\sigma^{\rm NNLO}_{gg \to H}$ that  is
obtained this way, using MSTW PDFs, is shown in Fig.~2. In the mass range $M_H
\! \approx \! 160$ GeV with almost the best sensitivity, one obtains  a 
$\approx +41\%,-37\%$ total uncertainty, to be compared to the $\approx\! 10\%$ 
and  $\approx\! 20\%$ uncertainties assumed, respectively, by the CDF and D0
collaborations. We also show for comparison, the result obtained when one adds
linearly, i.e.  as recommended by the LHC Higgs cross section working group, 
the  uncertainties from scale ($\!+\!20\%,\!-\!17\%$ on the sum of the jet cross
sections\footnote{An additional uncertainty of $\approx\! 7.5\%$ from jet
acceptance is introduced when considering the Higgs+jet cross sections.  We will
consider it to be experimental and, when added in quadrature to others, will
have little impact.} and PDFs ($+16\%,-15\%$ when the MSTW 68\%CL
PDF+$\Delta^{\rm exp} \alpha_s$  error is multiplied by a factor of two
following the PDF4LHC recommendation), leading to a total of $\approx \!+\!36\%,
\!-\!32\%$ for $M_H\! \approx\! 160$ GeV. Thus, the uncertainty that we assume
is comparable to the one obtained using the LHC procedure \cite{LHCXS}, the
difference being simply due to the  additional ${\cal O}(5\%)$ uncertainty from
the use of the EFT approach that we also include. \s

Let us stress again that the comparison between our values and those assumed by
the experimental collaborations becomes even worse when the cross section  is
evaluated with another set of PDFs.  For instance, with the HERA PDF
parametrization, there is a reduction of $\approx 40\%$ of the normalisation 
compared to the central value adopted in the CDF/D0 combined analysis. \s

Thus if the $\approx 20\%$ total uncertainty assumed by the CDF collaboration is
adopted, one can consider two scenarios. The first one is a reduction of
$\sigma^{\rm NNLO}_{gg \to H}$ by $\approx 20\%$ to account for the difference
between  the quadratic and (almost) linear ways of combining the individual
uncertainties.  A second scenario, would be simply to adopt the normalisation
obtained using the HERA PDFs which gives a $\approx 40\%$ reduction of
$\sigma^{\rm NNLO}_{gg \to H}$. In both cases, the remaining $\approx 20\%$
uncertainty due to scale variation and the EFT will correspond to the overall
theoretical uncertainty that has been assumed in the Tevatron analysis. \s

So far, we have only addressed the issue of the signal rate. However, it is 
clear that one should equally consider the same uncertainties in the background
cross sections. The by far largest background is $p\bar p\! \to\! W^+W^-$ for
which  CDF/D0 assume the inclusive cross section to be $\sigma  \!=\!  11.34
^{+4.9\%}_{-4.3\%} ({\rm scale})^{+3.1\%}_{-2.5\%} ({\rm PDF})$ pb. We have
reevaluated the rate using {\tt MCFM} \cite{MCFM} and find  $\sigma\!= \! 
11.55^{+5\%}_{-6\%} ({\rm scale})^{+5\%}_{-8\%} ({\rm 90\%CL\;PDF})$ pb using
the MSTW scheme (the errors due to $\alpha_s$ are negligible here) which gives
$\sigma= 11.55^ {+11\%}_{-14\%}$ pb if the errors are added according to
Ref.~\cite{Hpaper0}. In fact, if we adopt the ABKM or HERAPDF sets, we would
obtain  a rate of, respectively, 12.35 pb and 11.81 pb. i.e $\approx 9\%$ higher
in the maximal case. We will thus consider that  $\sigma(p\bar p\! \to\!
W^+W^-)$ can be $\approx 10\%$ larger/lower than assumed by CDF/D0\footnote{ We
have also evaluated $\sigma(p\bar p\! \to\! t\bar t)$  under the same
assumptions as \cite{Tevatron} but with  $m_t\!=\! 173.3\! \pm  \! 1.1$GeV and 
find  $\sigma(p\bar p \to t\bar t)\!= \!7.07 ^{+7.6\%}_{-8.6\%}{(\rm
scale)}^{+10.5\%}_{-8.0\%} ({\rm  PDF}\!+\!\Delta^{\rm exp+th}\alpha_s) {\pm
3.3\%} ( \Delta m_t)$ pb, which leads  using the procedure of \cite{Hpaper0}  to
a total uncertainty of  $\Delta \sigma/\sigma= ^{+15.6\%}_{-14.6\%}$, i.e  much
larger than the one assumed by CDF and D0.  In the case of the Drell--Yan
process, there is also a  $\approx 10\%$ excess in the rate if one uses HERAPDF
instead of  the MSTW set: $\sigma^{\rm HERA}_{ p\bar p\! \to \! Z} =7.6$ nb
versus  $\sigma^{\rm MSTW}_{p\bar p \! \to Z}=7$ nb \cite{PDF-all}.} and we will
consider a third  scenario in which the normalization of the  $p \bar p\! \to \!
WW$ background is changed by $\pm  10\%$. \s

Let us now come to the  discussion of the Higgs Tevatron exclusion limit in the
light of these theoretical uncertainties. We will base our exploration on the 
CDF study published in Ref.~\cite{CDF-analysis} which provides us with all the 
necessary details. In the analysis of the $gg \to H \to WW \to \ell \ell \nu \nu
$  signal, the production cross section has been broken into the three pieces
which yield different final state signal topologies, namely $\ell \ell \nu
\nu$+0\,jet,  $\ell\ell \nu \nu$+1\,jet and $\ell \ell \nu \nu$+2\,jets or more.
These channels which represent, respectively,  $\approx 60\%$, $\approx 30\%$
and  $\approx 10\%$ of the total $\sigma_{gg\to H}^{\rm NNLO}$ \cite{ADGSW}, 
have been studied  separately. In the $\ell \ell \nu \nu$+0\,jet and +1\,jet
samples, two configurations have been analyzed, one with a high and one with  a
low signal over background ratio (depending on the quality of the lepton
identification). In addition, a sample with a low invariant mass for the two
leptons, $M_{\ell \ell} \leq 16$ GeV, has been  included. Five additional
channels resulting from the contributions of the  Higgs--strahlung processes are
also included: $p\bar p \to VH \to VWW$ leading to same sign dilepton and to
trilepton final states. These channels give rather small signal rates, though.\s

Our main goal is to estimate the necessary relative variation of the integrated
luminosity needed to reproduce the currently quoted sensitivity of the CDF
collaboration,   if  the normalization of the Higgs signal cross section  (as
well as the corresponding backgrounds) is different from the one assumed to
obtain the results. \s

Our approach consists of the following.  First, we  try to reproduce as closely
as possible the CDF results using the information given in
Ref.~\cite{CDF-analysis}  for a mass $M_H\!=\!160$ GeV, for which the
sensitivity is almost the best (we will assume that the results  are similar in
the entire excluded mass range $M_H\approx 158$--175 GeV).  Then, we consider
the two scenarios discussed previously which, in practice, {\it reduce the
normalisation} of the Higgs production cross section by $\approx 20\%$ (when all
the  uncertainties are added  using the procedure of  Ref.~\cite{Hpaper0})  and
$\approx 40\%$ (when the HERAPDF set is used to derive the central value of the
cross section). We estimate the {\it relative variation} of the sensitivity  and
increase the integrated luminosity until we recover our initial sensitivity. 
Finally, we assume that the obtained relative variations of the sensitivity as
well as the  required luminosity to reproduce the initial sensitivity, would be
the same for the CDF experiment. \s

A naive attempt to reproduce the CDF results  \cite{CDF-analysis} was to use the
background, signal and data numbers for all the search channels of Tables
I--VIII   without including the neural-network  information or any treatment 
that uses shape information. This naive approach resulted in a sensitivity (95\%
CL/$\sigma_{\rm SM}$)  $\approx 12$ times weaker than the CDF one. This large
difference made us feel uncomfortable, as we would have needed to make the above
assumptions over one order of magnitude difference for the sensitivity (or two
orders of  magnitude for the resulting necessary  luminosity) compared to the
CDF analysis\footnote{This factor $\approx 12$ gain in sensitivity obtained
using neural network techniques (including spin--correlations, the main
discriminant), is to be compared with the modest gain of $\approx 30$--50\%
envisaged  by the LHC experiments. It turned, though, that the CDF cut--based
analysis was not  fully optimised. }. \s

To be as close as possible to the CDF analysis and results \cite{CDF-analysis},
we considered their neural network outputs for the 10 search channels  (each one
for the  signals, backgrounds and data) presented in Figs.~2,4,$\cdots$,16  to
build the background only and the background plus signal hypotheses, implemented
them in the  program {\tt MClimit} \cite{mclimit} and used a ratio of
log--likelihood ``\`a la LEP"  as a test--statistic for which we combined the
above channels; this provided the  95\%CL/$\sigma_{\rm SM}$  sensitivity limit
on the Higgs boson at the considered mass of $M_H\!=\!160$ GeV.  A median
expected 95\%CL/$\sigma_{\rm SM}$  limit of $S_0\!=\!1.35$ has been  obtained,
to be compared to $S_0\!=\!1.05$ in the CDF analysis; for the observed
95\%CL/$\sigma_{\rm SM}$  limit, the agreement is better as we obtain 1.35 
compared to 1.32. We feel thus satisfied with this  rather close result as  even
the CDF and D0 collaborations agree in their methods within only 10\% accuracy
for the same input Monte Carlo and data \cite{Talk-Tevatron}. We therefore
believe that we can  safely adopt the three working hypotheses described
above.\s 

\begin{figure}[!h]
\vspace*{-.6cm}
\centerline{\epsfig{file=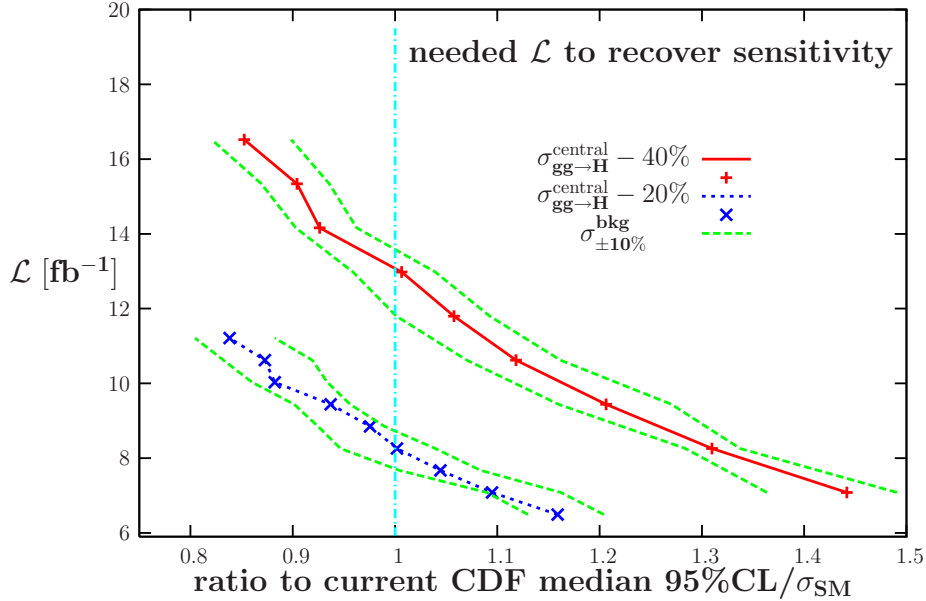,scale=0.7} }
\vspace*{-12.1cm}
\caption[]{The luminosity needed by the CDF experiment to recover the current 
sensitivity (with 5.9 fb$^{-1}$ data) when the $gg\!\to\! H\! \to \!\ell \ell 
\nu \nu$ signal rate is lowered by 20 and 40\% and with a $\pm 10\%$ change in 
the $p\bar p\! \to\! WW$  dominant background.}
\vspace*{-5mm}
\label{Lumi}
\end{figure}

We consider the first two scenarios in which the $gg\! \to\!  H\! \to\!  WW \!
\to \! \ell \ell \nu \nu$  signal cross section has been reduced by 20\% and
40\%.  In each case, the  expected signals and the corresponding  backgrounds at
the Tevatron  have been multiplied by a luminosity factor that has been varied. 
For each value of the  luminosity factor, the corresponding   median expected
95\%CL/$\sigma_{\rm SM}$ has been  estimated and normalized to the initial
sensitivity $S_0=1.35$ obtained above.  The results are reported in
Fig.~\ref{Lumi} where the Tevatron luminosity is shown  as a function of the
obtained normalised sensitivity.  The luminosity needed to recover the current
$S_0$ CDF sensitivity is given by the intersection of the vertical (blue) line
with the luminosity curves.  One sees that if $\sigma_{gg\to H}^{\rm NNLO}$ is
lowered by $20\%$, a luminosity of $\approx 8$ fb$^{-1}$, compared to  5.9
fb$^{-1}$ used in \cite{CDF-analysis} would be required for the same analysis to
obtain the current sensitivity. If the rate is lower by 40\%, the required
luminosity should increase to $\approx\! 13$ fb$^{-1}$, i.e. more  than a factor
of two, to obtain the present CDF sensitivity. \s

As an additional exercise, we also analyzed the impact of changing the
normalization of the background cross sections by $\pm 10\%$, as in our third
scenario, simultaneously with lowering the signal\footnote{The correlation
between signal and background is implici\-tly taken into account as we use the
results of \cite{CDF-analysis}; we assume though that it is almost the same when
another PDF set is adopted for both signal and background.} by 20 and 40\%. One
sees that increasing/decreasing the background will degrade/improve the
sensitivity  and a $\approx 10\%$ higher/lower luminosity would be required to
recover the sensitivity.\s

We conclude by noting that the reduction of the signal by 40\%  as would be the
case if the HERAPDFs were used for its normalization, would reopen the entire
mass range $M_H\!=\!158$--175 GeV excluded by the CDF/D0 analysis with 12.6
fb$^{-1}$ combined data. Hence, we face the uncomfortable situation  in which
the Higgs exclusion limit depends on the considered PDF.\smallskip 

{\bf Acknowledgements}: We thank G. Altarelli, M. Chen, A. Cooper-Sarkar, M.
Dittmar,  A. Korytov, H. Prosper, G. Salam, M. Spira, P. Verdier for
discussions. We acknowledge the projects SR/S2/JCB64 DST (India)
and ANR CPV-LFV-LHC NT09-508531 (FR). \bigskip

\section*{Erratum}

\begin{figure}[!h]
\begin{center}
\vspace*{1mm}
\epsfig{file=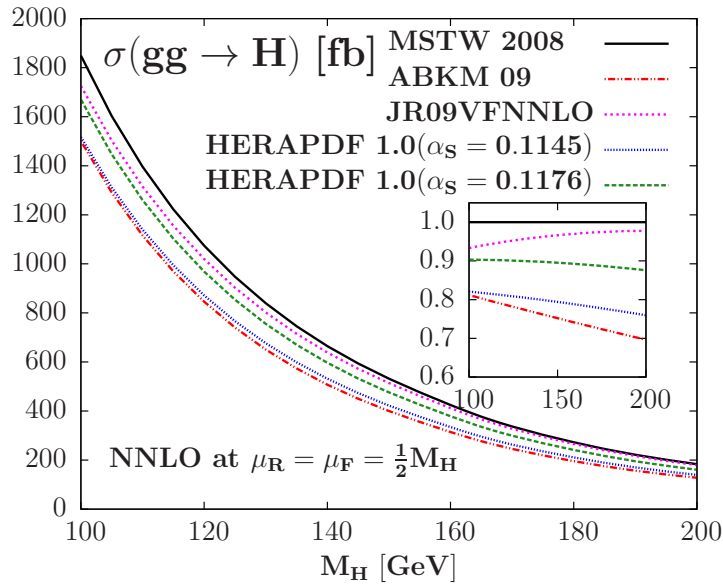,scale=0.89} 
\end{center}
\vspace*{-5mm}
\caption[]{The $gg\to H$ cross section as a function of $M_H$ when the four 
NNLO PDF sets, MSTW, ABKM, JR and HERAPDF, are used. In the inserts, shown
are the deviations with respect to the central MSTW value.}
\vspace*{-2mm}
\label{PDFs-correct}
\end{figure}

After our paper had appeared in Physics Letters B, we realised that an error
occurred in the numerical analysis which had led to Fig.~\ref{PDFs} for the
$gg\to H$ production cross section when the four NNLO PDF sets are adopted. In
the plot  with  the  two HERAPDF sets, the central scales at which $\sigma^{\rm
NNLO}_{gg\to H}$ has been evaluated were not set to $\mu_R=\mu_F=\frac12 M_H$ as
it should  have been, but at $\mu_R=\mu_F=\frac32 M_H$ which gives the minimal cross
section once the scale uncertainty is included. This explains the large
difference in the cross  section\footnote{We thank Graham Watt for pointing out
to us that his calculation of the $gg\to H$  cross section with HERAPDF  does
not lead to such a large difference.}, up to 40\%, between the MSTW and HERAPDF
predictions.  We thus present our  {\it mea culpa} and produce in
 Fig.~\ref{PDFs-correct} the correct  figure where all scales are 
consistently set to $\mu_R=\mu_F=\frac12 M_H$. The difference between the MSTW
and HERAPDF  predictions reduces now to $\approx 20\%$ at most, which is indeed
much more reasonable. In this case, the smallest value of the cross section is
given  when using the ABKM set and amounts to $\approx 20\%$--30\% in the
considered Higgs  mass range as noticed in Ref.~\cite{Hpaper0} (this difference
is slightly larger if the new ABM10 PDF set  is used \cite{PDF-all}). Note that
the same analysis presented for the LHC in  Ref.~\cite{Note-BR} is not affected
by this problem. \s

This error does not affect the subsequent discussion and almost does not change
our conclusions. Indeed, the main analysis which led to Fig.~2 (which, we
believe,  is the most important result of our paper) is still valid as we
estimate the PDF uncertainties within the MSTW set and our conclusion, that  the
theoretical uncertainty on the $gg\to H$  cross section at the Tevatron is
$\approx 40\%$, still holds true. \s

\begin{figure}[!h]
\vspace*{-.6cm}
\centerline{\epsfig{file=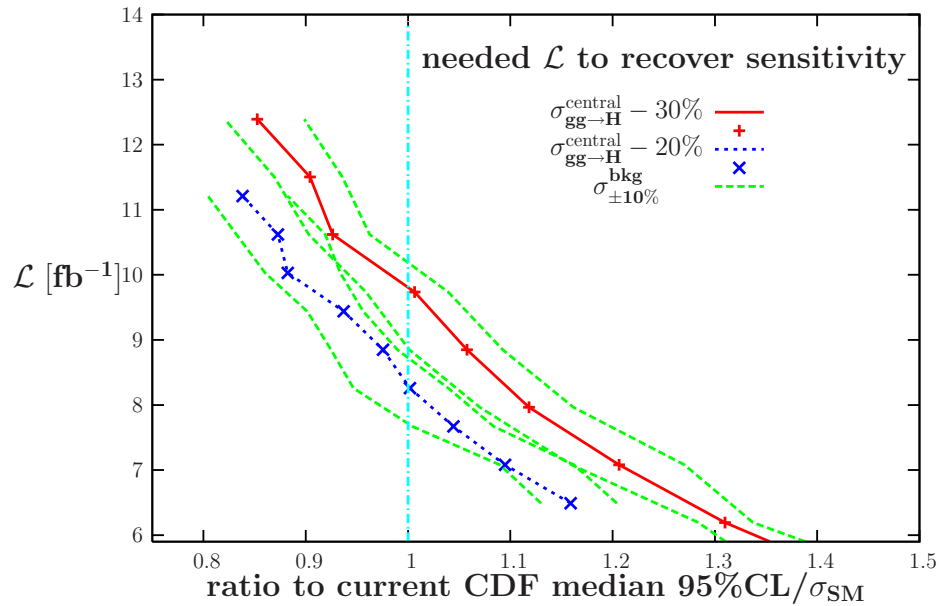,scale=0.7} }
\vspace*{-12.1cm}
\caption[]{The luminosity needed by the CDF experiment to recover the current 
sensitivity (with 5.9 fb$^{-1}$ data) when the $gg\!\to\! H\! \to \!\ell \ell 
\nu \nu$ signal rate is lowered by 20 and 30\% and with a $\pm 10\%$ change in 
the $p\bar p\! \to\! WW$  dominant background.}
\vspace*{-1mm}
\label{Lumi-new}
\end{figure}

Nevertheless, the interpretation of  the CDF/D0 limit when lowering the
normalisation of the cross section, has to be modified. Instead of lowering the
normalisation  by 40\%, one has to lower it by 30\% which is the difference
betweeen the  MSTW and ABKM  predictions. The luminosity needed by the CDF
experiment to recover the present sensitivity  is shown in Fig.~\ref{Lumi-new}
in this case. With this normalisation and  including the 10\% uncertainty on the
backgroud rate, the needed luminosity to recover the present sensitivity will be
slightly less than a factor of two. \s

Note, however, that the updated results given by the CDF/D0 experiments for the
winter 2011 conferences with a luminosity of 7.1 fb$^{-1}$ for CDF, lead to an
exclusion limit that is slightly worse than the one quoted here and only the
range $M_H=158$--173 GeV is excluded. Thus, even for a 30\% reduction of the
production cross section only instead of the 40\% used earlier, one still needs
$\approx 13$ fb$^{-1}$ data to recover the sensitivity obtained with 7.1
fb$^{-1}$.

\end{document}